\documentclass[final,3p,times,twocolumn]{elsarticle}
\usepackage{graphicx}
\usepackage{amssymb}
\usepackage{amsmath}
\usepackage{hyperref}
\usepackage{gensymb}

\usepackage{amsmath}
\usepackage{makeidx}
 
 \journal{Chaos,   Solitons and Fractals}

\usepackage[usenames]{color}
\usepackage[normalem]{ulem}
\usepackage{gensymb}

\usepackage{hyperref}
\usepackage[T1]{fontenc} 
\begin{document}

\title{Hollow cylindrical droplets in a very strongly dipolar condensate}


 \author{S. K. Adhikari}
\ead{sk.adhikari@unesp.br}

\address{Instituto de F\'{\i}sica Te\'orica, UNESP - Universidade Estadual Paulista, 01.140-070 S\~ao Paulo, S\~ao Paulo, Brazil}

\begin{abstract}

{ 

A harmonically trapped Bose-Einstein condensate (BEC) leads to topologically trivial compact states. 
Because of the long-range nonlocal dipole-dipole interaction, 
a strongly dipolar BEC  revealed many novel phenomena. 
Here we show    that in a strongly dipolar BEC  one  can have    a  hollow cylindrical quasi-one-dimensional metastable droplet  with ring topology while the  system is trapped only in the $x$-$y$ plane by a harmonic potential  and   
 a Gaussian hill potential  at the center and  untrapped  along the polarization $z$  axis.   In this numerical investigation 
we use  the imaginary-time propagation  of a     mean-field model where we include  the Lee-Huang-Yang interaction, suitably modified for dipolar systems.  Being metastable, these droplets are weakly stable  and we use real-time propagation to investigate its dynamics and establish stability.

}

\end{abstract}


\maketitle

\section{Introduction}

The size of  a usual harmonically-trapped Bose-Einstein condensate (BEC) has the same order of magnitude 
as the oscillator length.  
In a pioneering experiment \cite{y13,2d3} with 
 a strongly dipolar  harmonically-trapped  BEC of  $^{164}$Dy atoms  it was found that, with the increase of the number of atoms,  the system collapses to form a  stable tiny droplet of size much smaller than the oscillator lengths.  With the increase of number  of atoms, multiple droplets arranged on a lattice is formed.
Later on  it was demonstrated that  a single-droplet state can be stabilized even in  free space
under appropriate conditions \cite{santos,drop3,blakie}.   Quasi-one-dimensional (quasi-1D) droplets were studied in dipolar BECs under different conditions
\cite{c30,c31,c32,c33,c27}.   However, these droplets are compact objects.

More recently, in a remarkable experiment, Bigagli et al. observed and studied \cite{NaCs}
a very strongly dipolar BEC of NaCs molecules. In this system  the dipole-dipole interaction was  an order of magnitude larger than that in the   BEC of  dysprosium atoms.
Employing the microwave-shielding technique \cite{Bohn},  the effective electrical dipole moment of a NaCs molecule could be made very large so that the dipolar length 
      $a_{\mathrm{dd}} $  lies in  the range  $ 1000a_0$  to  $ 25000a_0$ \cite{NaCs} with 
       $a_{\mathrm{dd}}({\mathrm{NaCs}}) \gg a_{\mathrm{dd}}({\mathrm{Dy}}).$  
       Due to the very different nature of dipole-dipole interaction,
       it is expected that a very strongly dipolar  molecular BEC  
 may host different types of stationary  eigenstates, including those of the  droplets,   not possible in an atomic dipolar BEC.   
Because of the nonlocal, long-range  nature of  strong dipole-dipole interaction, this system  seems to be  ideal for studying different types  of novel droplets not possible in an atomic dipolar BEC.

Quantum  states of distinct  topology are of great general interest \cite{nat-gen} and such states can be created in a BEC by manipulating the confining trap so that the trap has a desired topology \cite{ska}. For example, a bubble-shaped or a ring-shaped  trap creates a BEC  of the same topology.  
  In a harmonically trapped nondipolar BEC,   only topologically trivial compact states can be obtained.  However, we demonstrated that   it is possible to have stable stationary  hollow cylindrical 
 eigenstates of ring topology \cite{luis13}, in addition to the droplet states \cite{y15,y16,y17,y18,y19}, 
 in a strongly dipolar  harmonically-trapped BEC of dysprosium atoms. 
 
 In this paper we demonstrate the formation of metastable  quasi-1D hollow cylindrical droplets of ring topology   in a very strongly dipolar  molecular BEC, appropriately trapped only  in the $x$-$y$ plane; the  ground state of this system is a stable  quasi-1D droplet.  
The trap in the $x$-$y$ plane is a circularly-symmetric harmonic potential plus a Gaussian hill potential at the center.   It was demonstrated in a theoretical study \cite{ska} that  it is possible to generate a condensate with a ring or bubble topology   in a trap with the same topology.  However,    the generation of a droplet with a distinct topology in this fashion was never investigated before.

In  the usual  mean-field  Gross-Pitaevskii (GP) model, including a nonlocal dipole-dipole interaction, if the dipole-dipole interaction  is larger than  a critical value, 
a    dipolar BEC collapses in 3D  \cite{c2,y6,c4,c5,c6}.
However, if we include a  strongly repulsive    Lee-Huang-Yang \cite{lhy} (LHY) interaction in this model, suitably modified  for a dipolar system \cite{qf1,qf2,qf3},  a stable stationary state can be obtained \cite{santos,drop3}.  The higher-order quartic  nonlinearity of the LHY interaction in  this improved mean-field model 
 stops the collapse, whereas  the attractive cubic nonlinearity of the nonlinear interactions in the 
  GP equation is responsible for providing the necessary binding.   
As the number $N$ of atoms (or molecules)   is increased in a strongly dipolar BEC, the condensate
 shrinks to a very small size due to the increasing     dipole-dipole attraction and with the exclusion of the collapse,  a stable dipolar droplet  can be formed \cite{y13,2d3,blakie}.  The size of a droplet is much smaller than the harmonic oscillator trap length in respective directions.
   Such a dipolar droplet can accommodate a critical maximum number of atoms  (or molecules) \cite{drop3}.   For a stronger dipole-dipole interaction, the formation of multiple-droplet states  \cite{blakie1,luis23,santos1,pohl}
 and hollow cylindrical states \cite{luis13} was demonstrated in the    improved mean-field model.   
 Recent experiments also confirmed \cite{y15,y16,y17,y18,y19} 
 the formation 
of multiple droplets     \cite{rev,rev2,rev3} in a strongly dipolar quasi-two-dimensional BEC.   

 These quasi-1D droplets are asymptotically free in the polarization $z$ direction and hence exhibit a soliton-like behavior. These droplets can move with a constant velocity along the $z$ direction as will be demonstrated. However,  technically a soliton  should be bound by  a balance between the dispersive repulsion and nonlinear attrction. The present droplets, in addition, are also subject to a higher-order repulsive LHY interaction. This is why we call these objects droplets, although the name soliton does not seem to be inappropriate.

We employ the improved mean-field model in  this study of hollow cylindrical quasi-1D metastable droplets with ring topology  in a very strongly  dipolar molecular BEC polarized along the $z$ direction.   
  In addition to the stable  one-droplet state \cite{drop3}, this system also has a metastable two-droplet state as the first excited state, whereas the hollow cylindrical  droplet is the second excited state.  We can obtain these metastable droplets by the imaginary-time propagation approach,  
 provided an initial state of appropriate symmetry is used.  To obtain an axially-symmetric hollow cylindrical droplet, 
 an axially-symmetric Gaussian initial state should be used.  The Gaussian hill potential at the center 
 allows to create the hollow central region of the droplet. To find a one-droplet (two-droplet) state easily in numerical simulation, one should use a one-droplet (two-droplet) initial state.
  We could not find any other type of metastable droplet states, such as a three-droplet or a four-droplet state. If the net dipole-dipole interaction of the system is reduced, by reducing either 
  the effective electric dipole moment of a molecule or 
  the number of molecules, only  a one-droplet ground state  can be obtained; the metastable two-droplet state and hollow cylindrical droplets cease to exist.  A very strong dipole-dipole interaction is essential for the formation of a  hollow cylindrical droplet.
   The metastable hollow cylindrical droplet and  the two-droplet state  are weakly stable and we study this dynamics by real-time propagation  of the   present  mean-field model.  For this purpose first we obtain the converged imaginary-time wave function for a specific hollow cylindrical droplet. Then we use  this solution as the initial state in a real-time propagation calculation after slightly modifying the angular frequency of the harmonic oscillator trap. A long-time radial oscillation of the droplet in the $x$-$y$ plane demonstrates its dynamical  stability.

    In Sec. \ref{II} a brief account of the  the improved mean-field model,   including a proper LHY interaction
    for a   dipolar BEC,  is given.     
      In Sec. \ref{III}   numerical  results  of this investigation for a strongly dipolar molecular BEC is presented.    The energy per molecule of the droplets is found to saturate with the increase of number of molecules.  We present results for the  3D isodensity profile of the droplets in addition to the  integrated 
      1D density along the polarization $z$ direction
       and the integrated 
      2D \cite{ajp}
      density in the $x$-$y$ plane.
We generate the stationary states of the metastable droplets by        imaginary-time propagation 
  and study the  dynamics  by real-time propagation.  We elaborate  a  brief account  of the principal findings of this  investigation in Sec. \ref{IV}.

\section{Improved Mean-field model}

\label{II}

At ultra-low temperatures the properties of a very strongly dipolar condensate of $N$ molecules,
polarized along the $z$ axis,  each of
mass $m$, can be described by the following improved mean-field GP equation  with a nonlocal nonlinearity, including the appropriate \cite{qf1,qf2} LHY interaction \cite{lhy}, for the 3D wave function $\psi({\bf  r}, t)$:   \cite{santos1,dip,yuka,dipbec,dipbecx}
\begin{align}\label{eq.GP3d}
 \mbox i \hbar \frac{\partial \psi({\bf r},t)}{\partial t} &=\
{\Big [}  -\frac{\hbar^2}{2m}\nabla^2
+\frac{1}{2}{m\omega^2}(x^2+y^2) 
+ V_0 e^{-\frac{x^2+y^2}{\delta^2}} 
\nonumber\\                                                                
&+ \frac{4\pi \hbar^2}{m}{a} N \vert \psi({\bf r},t) \vert^2 \nonumber \end{align}
\begin{align}
& +\frac{3\hbar^2}{m}a_{\mathrm{dd}} 
N\int\frac{1-3\cos^2 \theta}{|{\bf  R}|^3} 
\vert\psi({\mathbf r'},t)\vert^2 d{\mathbf r}'\nonumber\\
& +\frac{\gamma_{\mathrm{LHY}}\hbar^2}{m}N^{3/2}
|\psi({\mathbf r},t)|^3
\Big]  d {\bf r } \, ,
\end{align}
which we use for studying the statics and dynamics of a hollow cylindrical droplet. 
Here the trapping potential consists of a Gaussian hill potential at the center and 
a weak axially-symmetric  harmonic trap with no confinement along the $z$ direction, $\delta$ and $V_0$ are the  width and the
strength parameters of the Gaussian hill and the angular frequency of the  harmonic confinement is $\omega$. 
Here, ${\bf r} \equiv \{x,y,z\}$ and ${\bf r'} \equiv \{x',y',z'\}$ are the positions  of two  dipolar molecules 
and $\theta$ is the angle between   the relative vector  $\bf R\equiv r-r'$ and  the  polarization
$z$ axis.   
 The axially-symmetric repulsive Gaussian potential in Eq. (\ref{eq.GP3d}) is responsible for the formation of  the metastable hollow cylindrical  droplet with long extension in the $z$ direction, whereas the strong dipole-dipole repulsion in the $x$-$y$ plane stabilizes this droplet. 
  The normalization of the wave function  is $\int \vert \psi({\bf r},t) \vert^2 d{\bf r}=1.$ 
  The dipolar length $a_{\mathrm{dd}}$  in  Eq. (\ref{eq.GP3d}), is given by \cite{NaCs}
\begin{align}\label{eq.dl}
a_{\mathrm{dd}} =\frac{m \mathrm{d_{eff}}^2 }{ 12\pi \hbar^2 \epsilon_0},
\end{align}
and measures the strength of the dipole-dipole interaction, whereas  the scattering length $a$ measures the  strength of the contact interaction. In Eq. (\ref{eq.dl}) $\epsilon_0$ is the
permittivity of vacuum and $\mathrm{d_{eff}}$ is the effective electric dipole moment of a molecule.

 The  LHY interaction  coefficient   $\gamma_{\mathrm{LHY}}$ in Eq. (\ref{eq.GP3d})
is  \cite{qf1,qf2}
\begin{align}\label{qf}
\gamma_{\mathrm{LHY}}= \frac{128\sqrt{\pi a^5} }{3}Q_5(\varepsilon_{\mathrm{dd}}),
\end{align}
where  $ \varepsilon_{\mathrm{dd}} ={a_{\mathrm{dd}}}/{a}$ and the auxiliary function   
\begin{equation}
 Q_5(\kappa)\equiv  \int_0^1 d\xi(1+3\kappa\xi^2-\kappa)^{5/2} 
\end{equation}
can be written as \cite{blakie}
\begin{align}\label{exa} 
Q_5(\kappa) &=\
\frac{3^{3/2}\kappa^{5/2}}{16}  \Re \left[  
 15\eta^3 \mathrm{ln} \left( \frac{1+\sqrt{1+\eta}}{\sqrt{\eta}}\right)  \right.\nonumber\\
&\left. +(8+26\eta+33\eta^2)\sqrt{1+\eta}  \right], 
\end{align}
where  $\eta =  (1-\kappa)/(3\kappa)$ and  $\Re$ is the real part of the function.  We use expression (\ref{exa}) for the auxiliary function  $Q_5(\kappa)$
in our numerical study.

  It is convenient to write Eq. (\ref{eq.GP3d})  in dimensionless form by a transformation of the variables. 
  We scale lengths in units of $l_0 = \sqrt{\hbar/m\omega_0}$, where $\omega_0$ is a constant frequency, 
  angular frequency  $\omega$ in units of $\omega_0$,  potential $V_0$ and energy in units of $\hbar \omega_0$,
   $|\psi|^2$ in units of $l_0^{-3}$, time in units of $t_0=\omega_0^{-1}$ and obtain
\begin{align}\label{GP3d2}
\mbox i \frac{\partial \psi({\bf r},t)}{\partial t} & =
{\Big [}  -\frac{1}{2}\nabla^2 + \frac{1}{2}\omega^2\left(x^2+ y^2\right) +V_0 e^{-\frac{x^2+y^2}{\delta^2}    }\nonumber \\
&+4\pi{a} N \vert \psi({\bf r},t) \vert^2 
\nonumber\\ &
+3a_{\mathrm{dd}}  N
\int 
\frac{1-3\cos^2 \theta}{|{\bf  R}|^3} 
\vert\psi({\mathbf r'},t)\vert^2 d{\mathbf r}'   \nonumber \\ 
&+\gamma_{\mathrm{LHY}}N^{3/2}
|\psi({\mathbf r},t)|^3  
\Big] \psi({\bf r},t).
\end{align}
In this equation and in the following we use the same symbols to denote the transformed and the original variables without any risk of confusion.
The normalization of the transformed wave function remains unchanged:  $\int d{\bf r} |\psi({\bf r,}t)|^2 =1.$

Equation   (\ref{GP3d2})  can also be derived      using  the variational principle 
\begin{align}
{\mathrm i} \frac{\partial \psi}{\partial t} = \frac{\delta E}{\delta \psi^*},
\end{align}
where the energy   functional $E$ is the energy per molecule of the  stationary droplets and is given  by \cite{rmp}
\begin{align} \label{energy}
E &= \int d{\bf r} \left[ {\frac{1}{2}|\nabla\psi({\bf r})|^2} \right. \nonumber \\
 &+ \left\{ \frac{1}{2}\omega^2\left(x^2+ y^2\right) +V_0 e^{-\frac{x^2+y^2}{\delta^2} }  \right\}
|\psi({\bf r})|^2 
\nonumber  \\
&+ \frac{3}{2}a_{\mathrm{dd}}N|\psi({\bf r})|^2 
\left. \int  
\frac{1-3\cos^2 \theta}{|{\bf  R}|^3} 
|\psi({\bf r'})|^2 d {\bf r'} \right. \nonumber \\
&\left. + 2\pi aN |\psi({\bf r})|^4 +\frac{2}{5}\gamma_{\mathrm{LHY}} N^{3/2}
|\psi({\bf r})|^5\right].
\end{align}

\section{Numerical Results} 

\label{III}
 
We can  obtain the very strongly dipolar  metastable droplets of  molecules 
by  solving  numerically the  improved mean-field    
 Eq.  (\ref{GP3d2}),   employing  available C/FORTRAN programs \cite{yuka1,dip} or their open-multiprocessing counterparts \cite{omp,ompF},
using  the split-time-step Crank-Nicolson  discretization rule  \cite{crank}. 
We employ  imaginary-time propagation for the stationary state of the droplet  and study its dynamics by real-time propagation. 
  The dipole-dipole potential is divergent at short distances and hence the numerical treatment of this potential requires some care. The integral 
over the dipole-dipole potential is evaluated in Fourier (momentum) space by a convolution identity \cite{dip,dipbec} requiring
the Fourier transformation of the dipole-dipole potential and
density. The Fourier transformation of the dipole-dipole potential can be analytically evaluated \cite{dip}. The remaining
Fourier transformations are evaluated numerically using
a fast Fourier transformation algorithm.
First we solve   the problem   in  the momentum space, then we employ a backward Fourier transformation to obtain the final
 solution in configuration space.
 In the Crank-Nicolson discretization we use space steps $dx=dy= 0.1,
dz=0.2$  and employ  in imaginary-time propagation time step 
$dt=0.1\times (dx \times dy \times dz)^{2/3}$  and in real-time propagation
$dt=0.025 \times (dx \times dy \times dz)^{2/3}$.    The number of space discretization points in $x$ and $y$ directions were $N_x=N_y=257$ and that in the $z$ direction could be as large as $N_z=1025$.

  In this study  of a quasi-1D hollow cylindrical droplet  and a 
  two-droplet state 
  we consider  a very strongly dipolar molecular  BEC  with 
  {$a_{\mathrm{dd}}\gg a$}.    Here  we will be employing  $a_{\mathrm{dd}} = 2000a_0$ and  $a=100a_0$, in order to have  a very strongly dipolar system appropriate for this investigation.  We consider the  NaCs molecules of mass    $m$(NaCs)    $\approx 156 \times 1.66054\times 10^{-27}$ kg; 
 we take the constant frequency scale  {$\omega_0 = 2\pi \times 260 $ } Hz, so that, the   length scale  $l_0 =\sqrt{\hbar/m\omega_0}= 0.500$  $\mu$m, time scale  {$t_0=\omega_0^{-1}=0.61$} ms.   The trap parameters in Eq. (\ref{eq.GP3d}) will be taken as
 {$\omega =2\pi \times 78$ Hz, $V_0/h = 2600$ Hz, $\delta = l_0= 0.5$ $\mu$m.} 
 The arbitrary choice of the reference frequency $\omega_0$ will have no effect on the final results of this investigation. 
{Although we are using the NaCs molecules as a reference, we cannot describe the actual experimental situation \cite{NaCs} of a dipolar BEC under microwave shielding. This is because, the presence
of electromagnetic or microwave fields, in microwave-shielded dipolar BEC, can modify the
long-range forces between ultracold dipolar molecules \cite{Bohn,58,dm1,dm2}
 in a way, which is not yet fully understood.
}

\begin{figure}[t!]  
\begin{center}
\includegraphics[width=\linewidth]{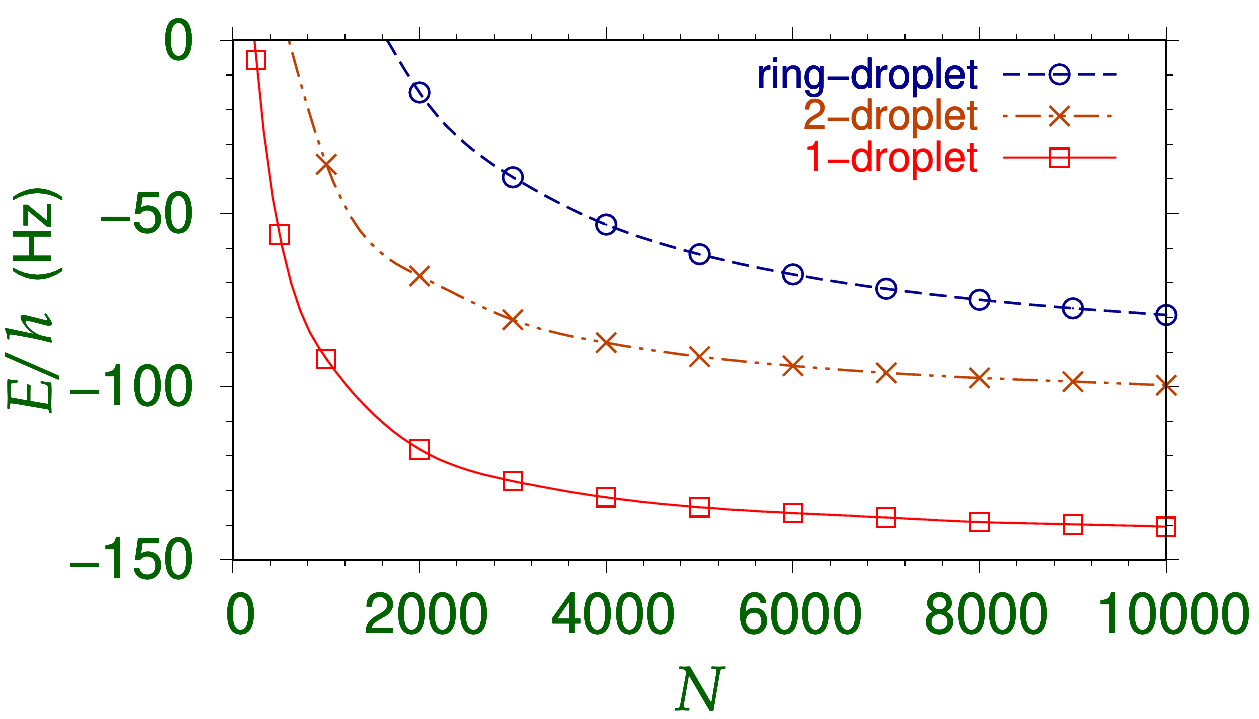}

\caption{The energy functional $E$, energy per molecule,   in units of the Planck constant $h$, e.g.  $E/h$ in Hz, of a hollow cylindrical ring topology droplet, two-droplet state and a one-droplet state  for different number 
of  molecules $N$. The numerical results (points) are joined smoothly by the lines. In Figs. \ref{fig1} to \ref{fig6}, the dipolar length and the scattering lengths  are $a_{\mathrm{dd}}=2000a_0, a=100a_0$, respectively, and    the trap parameters, viz. Eq. (\ref{eq.GP3d}), are {$\omega =2\pi \times 78$ Hz, $V_0/h = 2600$ Hz,  $\delta = 0.5$ $\mu$m and    $ l_0=0.5$ $\mu$m.  }}

\label{fig1}   
\end{center}
\end{figure}

  We can obtain a two-droplet state easily, employing the  imaginary-time propagation method, if  in the initial state the two droplets are placed symmetrically  at $\{x,y\}=\{0,\pm \alpha\}.$ The cylindrically-symmetric confining potential, viz. Eq. (\ref{eq.GP3d}), in the $x$-$y$ plane is one with ring topology. Hence,  to obtain a hollow cylindrical droplet, the initial state can simply be a cylindrically-symmetric Gaussian state.
The trap symmetry   naturally transforms the initial Gaussian state to a hollow cylindrical droplet.   
Because of the distinct spatial symmetry of these two types of metastable droplets,  these eigenstates are surprisingly stable and  a higher-energy   hollow cylindrical  metastable droplet does not easily transform  to a  two-droplet metastable droplet or a one-droplet ground state  with lower energy in an imaginary-time calculation,  provided we use  a numerical scheme with small space  and time  steps and  an appropriate  number of discretization points along the three axes.  However, in an approximate numerical calculational scheme employing imaginary-time propagation  generating
 larger numerical error
this will not be true and a
hollow cylindrical  state  with higher energy  may eventually transform  to a  lower-energy   two-, or,   a one-droplet state.

Because of the large molecular dipole-dipole interaction,  the present droplets can be realized for a relatively small number $N$ of molecules. We plot
 in Fig. \ref{fig1}  the energy  functional $E$  in units of Planck constant $h$, e.g.  $E/h$, for the hollow cylindrical droplet, and the  two-, and the one-droplet states versus the  number $N$ of  molecules ($N \le 10000$). 
  For a fixed number $N$ of molecules, the energy functional  $E/h$   
  of a hollow cylindrical droplet, which is the second excited state,  is larger than that of the two-droplet state, which is the first excited state.  The one-droplet state is the ground state.
  For each type of droplet, 
the energy $E/h$, and   the total energy of the system  $NE/h$,    
 decrease with the increase of the  number $N$  of molecules as we can find in Fig. \ref{fig1}.
 Nevertheless, the energy per molecule $E/h$  has a tendency to saturate for very large $N$ in this model.
  For the present set of parameters, in imaginary-time propagation we could not stabilize any other type of droplet, e.g. a three- or a four-droplet state. In imaginary-time propagation, 
such initial states  eventually transform to a  hollow cylindrical droplet or a   two-droplet or a one-droplet state. { The three types of droplet states shown in Fig. \ref{fig1} $-$ ring-droplet, two-droplet and one-droplet states $-$ coexist for the same set of parameters.}

\begin{figure}[t!]
\begin{center}
\includegraphics[width=\linewidth]{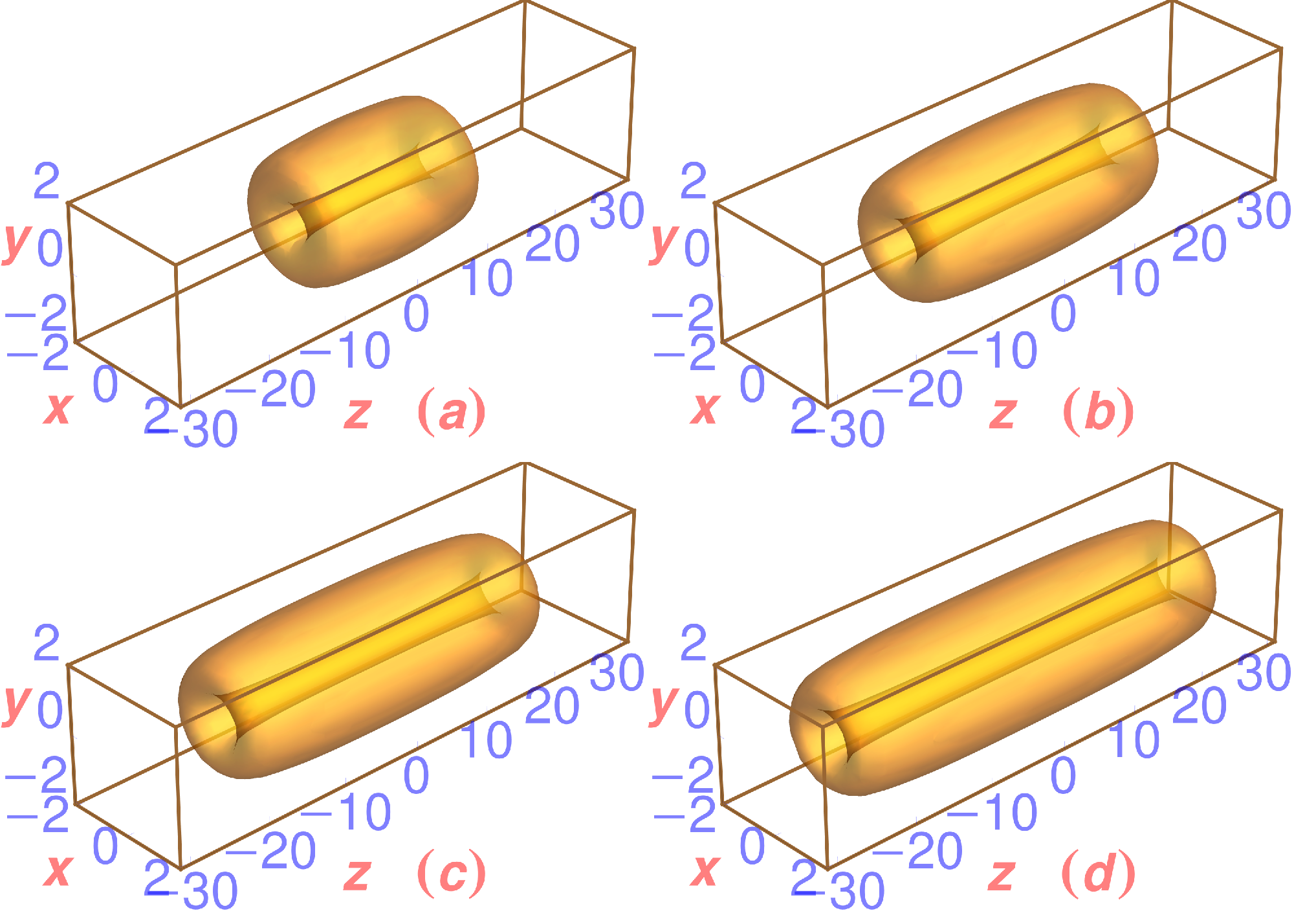}

\caption{Isodensity profile of   normalized density $|\psi(x,y,z)|^2$
 (normalization $\int d{\bf r}|\psi({\bf r})|^2 =1$) of a hollow cylindrical  droplet of $N=$ { (a) 2000, (b) 4000, 
(c) 6000, and (d) 8000}  molecules. 
The density on the contour is 
{$n_{\mathrm{contour}}=$} {$4 \times 10^{10}$} cm$^{-3}$.  The maximum density inside  these droplets are 
(a) $4 \times 10^{11}$ cm$^{-3}$,  (b) $2.5 \times 10^{11}$ cm$^{-3}$,   (c)  $2.3 \times 10^{11}$ cm$^{-3}$, and 
(d) $2 \times 10^{11}$ cm$^{-3}$. 
  As the net  molecular density  in the droplet is $N|\psi(x,y,z)|^2$, the real  molecular density  on the contour of these droplets is $N n_{\mathrm{contour}}$. The unit of length is $\mu$m.  All parameters are the same as in Fig. \ref{fig1}.
}

\label{fig2} 
\end{center}
\end{figure}
  
 We display in  Fig. \ref{fig2}  the isodensity profile of   normalized density $|\psi(x,y,z)|^2$ of  a hollow cylindrical  droplet   for (a) $N=2000$,
 (b) $N=4000,$  (c) $N=6000,$ and (d) $N=8000$ molecules, which is the principal finding of this paper. The central hollow region in the droplets  is clearly visible in these plots.  
  With the increase of the  number $N$ of molecules  from Fig. \ref{fig2}(a) through (d), the  net dipole-dipole interaction of the system increases  and the length of the droplets along the polarization $z$ 
  axis also increases. However  the spatial extension of the droplet  in the $x$-$y$ plane  
  remains essentially the same for different 
  number   $N$  of molecules in Figs. \ref{fig2}(a)-(d).  
 The outer radius of the   hollow cylindrical  droplet in the $x$-$y$ plane 
 is essentially controlled by the minimum of the confining trap, viz. (\ref{eq.GP3d}), in that plane.  Hence the radii of the four droplets in the same trap in Fig. \ref{fig2}  are approximately the same, although this radius in the $x$-$y$ plane reduces a little with the increase of $N$ from Fig. \ref{fig2}(a) to Fig. \ref{fig2}(d), due to the increase of the net dipole-dipole interaction.

\begin{figure}[t!]
\begin{center}
\includegraphics[width=\linewidth]{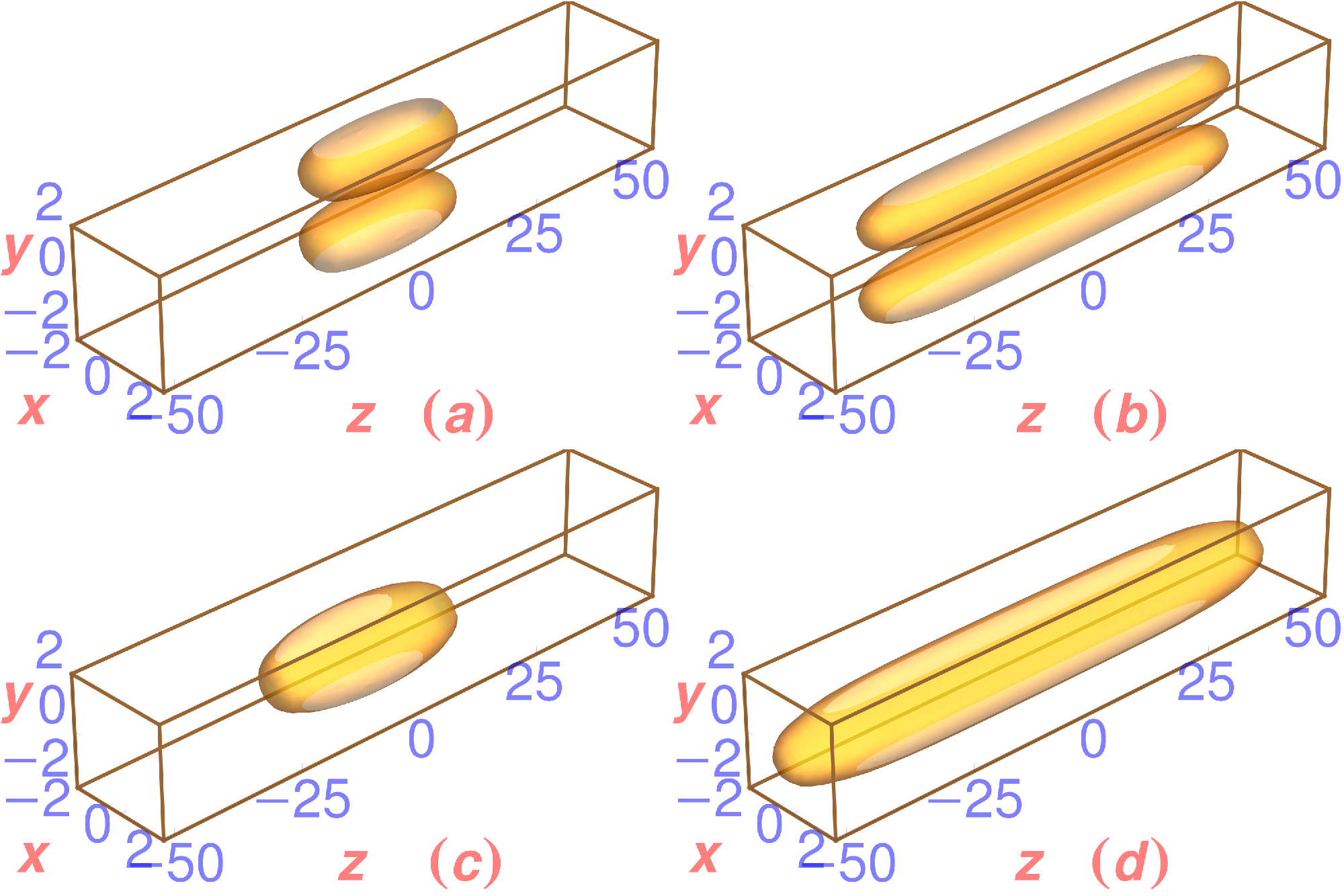} 

\caption{Isodensity profile of    normalized density $|\psi(x,y,z)|^2$  (normalization $\int d{\bf r}|\psi({\bf r})|^2 =1$) of a two-droplet state of  $N=$  (a) 2000 and  (b) 8000  molecules. The  same of a  one-droplet state of 
(c) 2000, and (d) 8000  molecules.  The density on the contour is 
{$n_{\mathrm{contour}}=$} {$4 \times 10^{10}$} cm$^{-3}$.  The unit of length is $\mu$m.
All parameters    are the same as in Fig. \ref{fig1}.
}  
\label{fig3} 
\end{center}
\end{figure}

{Although the hollow cylindrical states illustrated in Fig. \ref{fig2} bear some similarity with a vortex state,  these have zero angular momentum and are not vortex states. 
The possibility of the appearance of vortex droplets with a hollow cylindrical profile in a dipolar BEC was demonstrated in Ref. \cite{aa} and briefly discussed  in Ref. \cite{b}. In these cases the  vortex axis is parallel to the polarization direction of dipolar atoms. However, it was shown in Refs. \cite{c,d} that the vortex modes become more stable when the vortex axis is perpendicular to the polarization direction.
}

A metastable two-droplet state and a stable one-droplet state share many properties of a hollow cylindrical state. In Fig.  \ref{fig3} we present the isodensity profile of  normalized density $|\psi(x,y,z)|^2$  of a two-droplet state of (a) $N=2000$
 and (b) $N=8000$ molecules as well as a one-droplet state of (c) $N=2000$ and (d) $N=8000$  molecules.  
 For a fixed  number  of molecules $N$,  a  
 one-droplet state displayed  Figs. \ref{fig3}(c)-(d)  is longer along the $z$ axis than  a two-droplet state presented in Figs. \ref{fig3}(a)-(b), respectively, for $N=2000$ and 8000. This is because, for  a fixed $N$,  a one-droplet state has twice as many molecules when compared to  a two-droplet state.  A larger density of dipolar molecules in a one-droplet state leads to a larger dipole-dipole interaction which  generates a longer droplet.  The hollow cylindrical droplet of Fig. \ref{fig2} can be considered to be a continuous form of a multi-droplet droplet and, for a fixed $N$, this is shorter than a two- and a one-droplet state.

{ To study the distribution of matter in these droplets, it is  convenient to consider the following 
integrated 2D density $n_{\mathrm{2D}}(x,y)$ in the $x$-$y$ plane  and the integrated 1D density $n_{\mathrm{1D}}(z)$ along the $z$ axis   defined by \cite{ajp}
\begin{align}\label{2d}
n_{\mathrm{2D}}(x,y) &= \int_{-\infty}^{\infty}   |\psi(x,y,z)|^2  dz,\\
 n_{\mathrm{1D}}(z) &= \int_{-\infty}^{\infty}  \int_{-\infty}^{\infty}   |\psi(x,y,z)|^2 dx dy,
 \label{1d}
\end{align}
respectively.
The integrated 2D density (\ref{2d}) is useful to better visualize the localization of the droplet in the $x$-$y$ plane, and the integrated 1D density  (\ref{1d}) is useful to have a clear vision of  the linear extension of the  same along the $z$ axis.}

\begin{figure}[t!]
\begin{center}
 
\includegraphics[width=\linewidth]{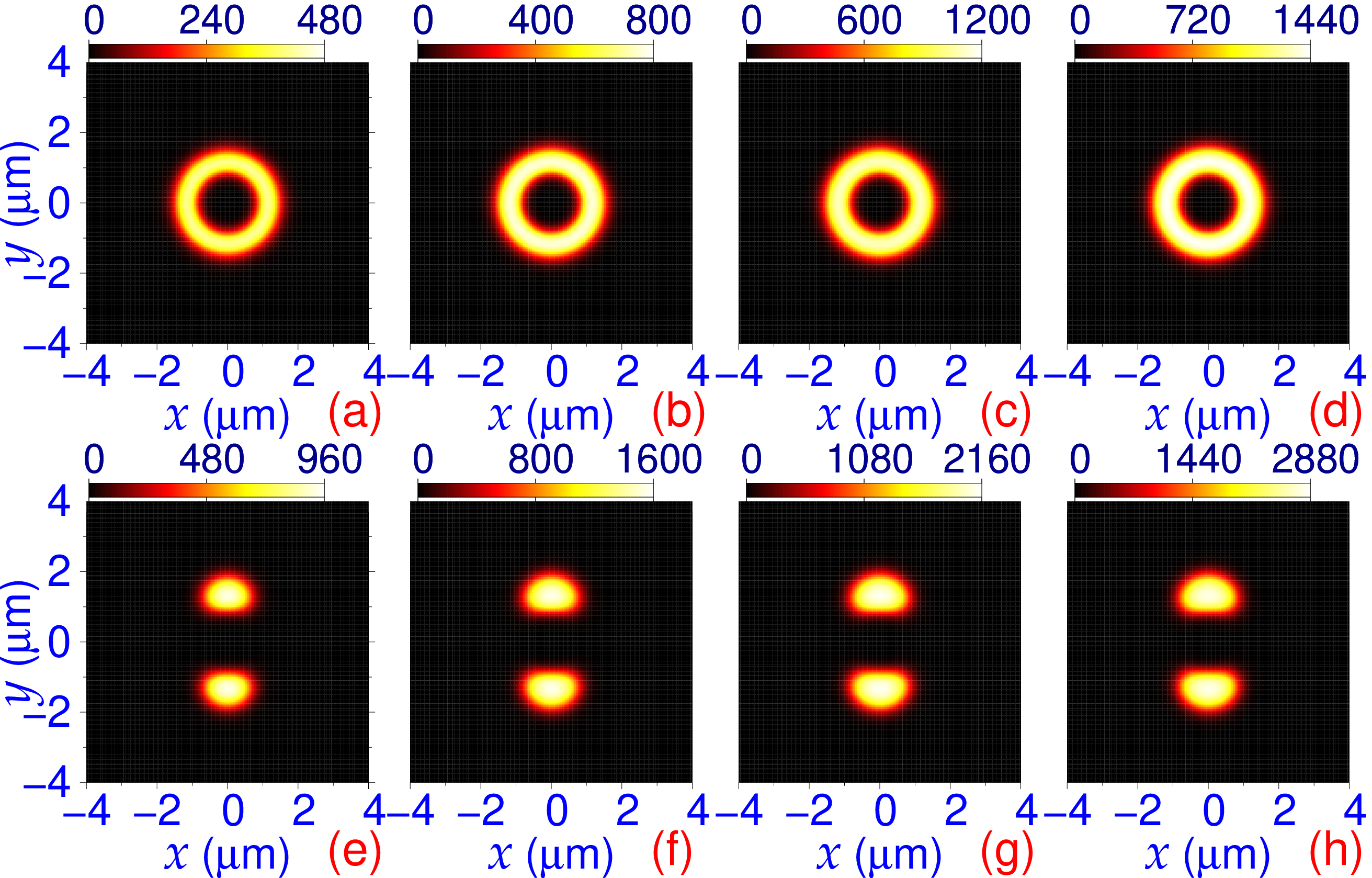} 

\caption{Contour   plot of  integrated net 2D density $n(x,y) \equiv Nn_{\mathrm{2D}}(x,y)$, viz. Eq. (\ref{2d}), 
[normalization $\int_{-\infty}^\infty dx \int_{-\infty}^\infty  dy n(x,y)=N$]
versus $\{x,y\}$ of hollow cylindrical  droplets of ring topology illustrated in  Fig. \ref{fig2} for  $N=$  {(a) 2000, (b) 4000, 
(c) 6000, and (d) 8000}  molecules.  The same  of two-droplet states illustrated in  Fig.  \ref{fig3} for  $N=$  {(e) 2000, (f) 4000, 
(g) 6000, (h) 8000} molecules.figure
The unit of densities in the color box is $\mu$m$^{-2}$.
All parameters are the same as in Fig. \ref{fig1}.
}

\label{fig5} 
\end{center}
\end{figure}

    {
In Fig. \ref{fig5} we display a contour plot of  the net integrated 2D density $n(x,y) \equiv Nn_{\mathrm{2D}}(x,y)$,  as a function of $x$ and $y$,  of the hollow cylindrical  droplets  of Fig. \ref{fig2}    of  $N=$  (a) 2000, (b) 4000, 
(c) 6000, and (d) 8000 molecules.
 The 2D density $n(x,y)$ 
 of the two-droplet states  of Fig. \ref{fig3}   are shown in Fig. \ref{fig5} for   $N=$  (e) 2000, (f) 4000, 
(g) 6000, and (h) 8000  molecules.  With the increase of 
 the number $N$ of molecules  in Figs. \ref{fig5}(a)-(d) of a hollow cylindrical  droplet,  these acquire a larger density of molecules explicitly shown  in the overhead color box.  The same trend can also be 
 found in the case of a two-droplet state illustrated in  Figs. \ref{fig5}(e)-(h).  
 Each droplet of a two-droplet state becomes fatter as $N$ increases, e.g. its spatial intersection in the $x$-$y$ plane increases.   If we compare the contour plots in Figs. \ref{fig5}(a)-(d) for a hollow cylindrical droplet
 with  the same in  Figs. \ref{fig5}(e)-(h) for a two-droplet state for      $N=$   2000,  4000, 6000, and  8000,  respectively,  we find that the total intersection of the droplet in the $x$-$y$ plane is larger in the former case. Consequently,  the density of molecules in the case of a two-droplet state is larger as can be seen in the corresponding  overhead color box.

}

 \begin{figure}[t!]
\begin{center}
\includegraphics[width=.49\linewidth]{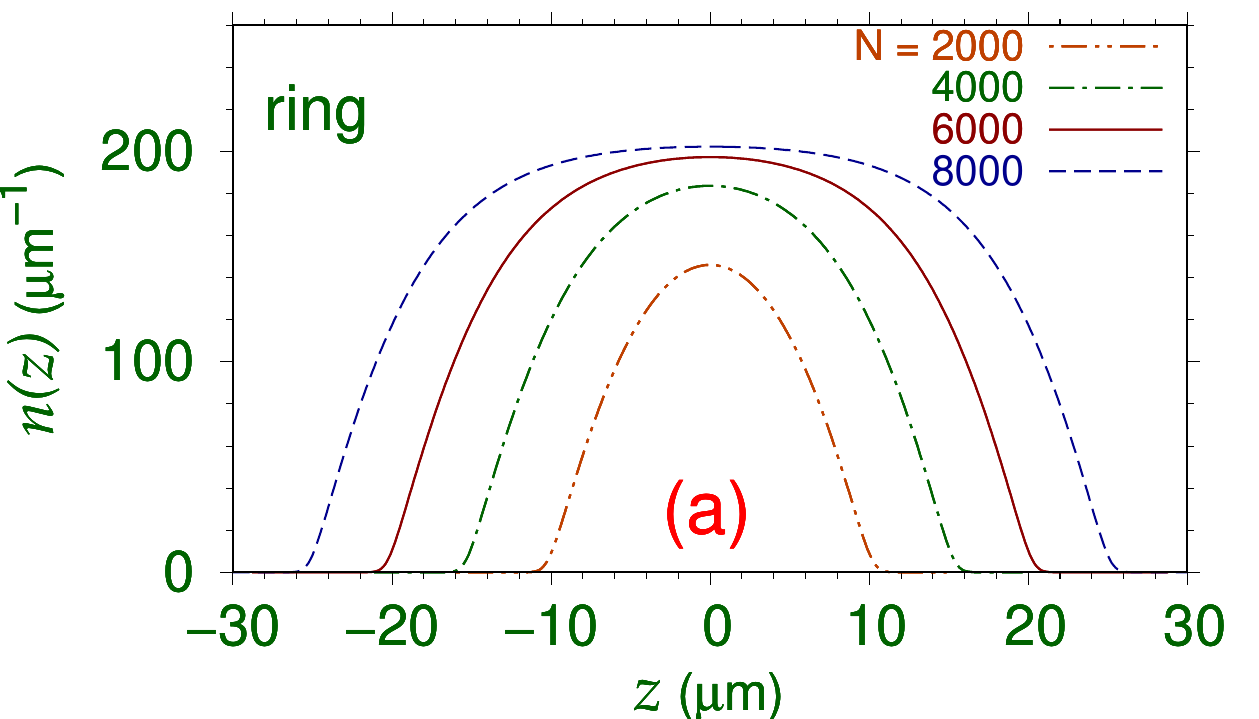} 
\includegraphics[width=.49\linewidth]{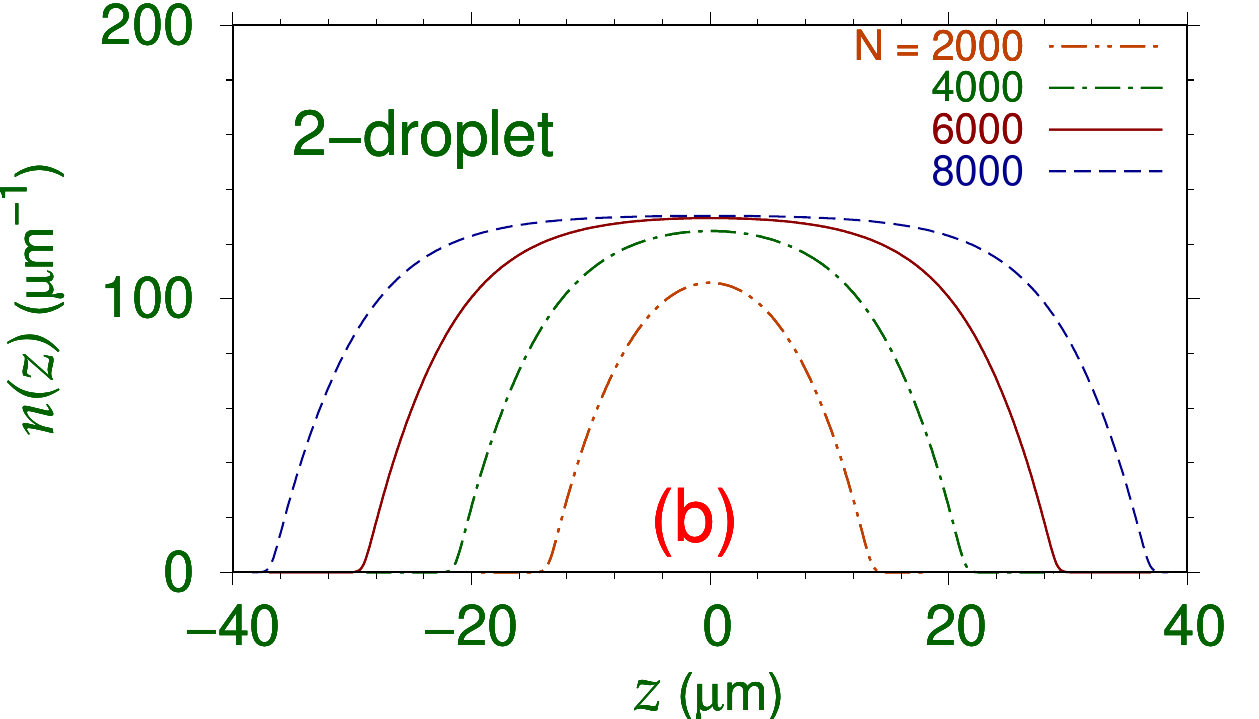}

\caption{Integrated net 1D density $n(z) \equiv Nn_{\mathrm{1D}}(z)$, viz. Eq. (\ref{1d}),  [normalization $\int_{-\infty}^\infty dz n(z)=N$] versus $z$ of (a) a hollow cylindrical droplet of ring topology  and  (b) a two-droplet  droplet  illustrated in  Figs. \ref{fig2} and \ref{fig3}, respectively, for different $N=2000,4000,6000$, and  8000  molecules. 
All parameters are the same as in Fig. \ref{fig1}. 
}

\label{fig6} 
\end{center}
\end{figure}  

 In Fig. \ref{fig6} we plot the net integrated 1D density   $n(z) \equiv N n_{\mathrm{1D}} (z)$
versus $z$ for (a) a hollow cylindrical droplet and (b) a
two-droplet state for $ N = 2000, 4000, 6000$ and $8000.$
From these plots we can directly see the linear extension of these droplets along 
the polarization $z$ direction. In both cases, as expected from  the isodensity profile of the droplets in Figs. \ref{fig2} and \ref{fig3}, the length of the droplet 
increases as the number of molecules increases. 
Also, the length of 
a two-droplet state  along the $z$ axis is larger than   that of a hollow cylindrical
droplet  for a fixed $N$. The one-droplet state is the longest of all the droplets

 \begin{figure}[t!]
\begin{center}
\includegraphics[width=.48\linewidth]{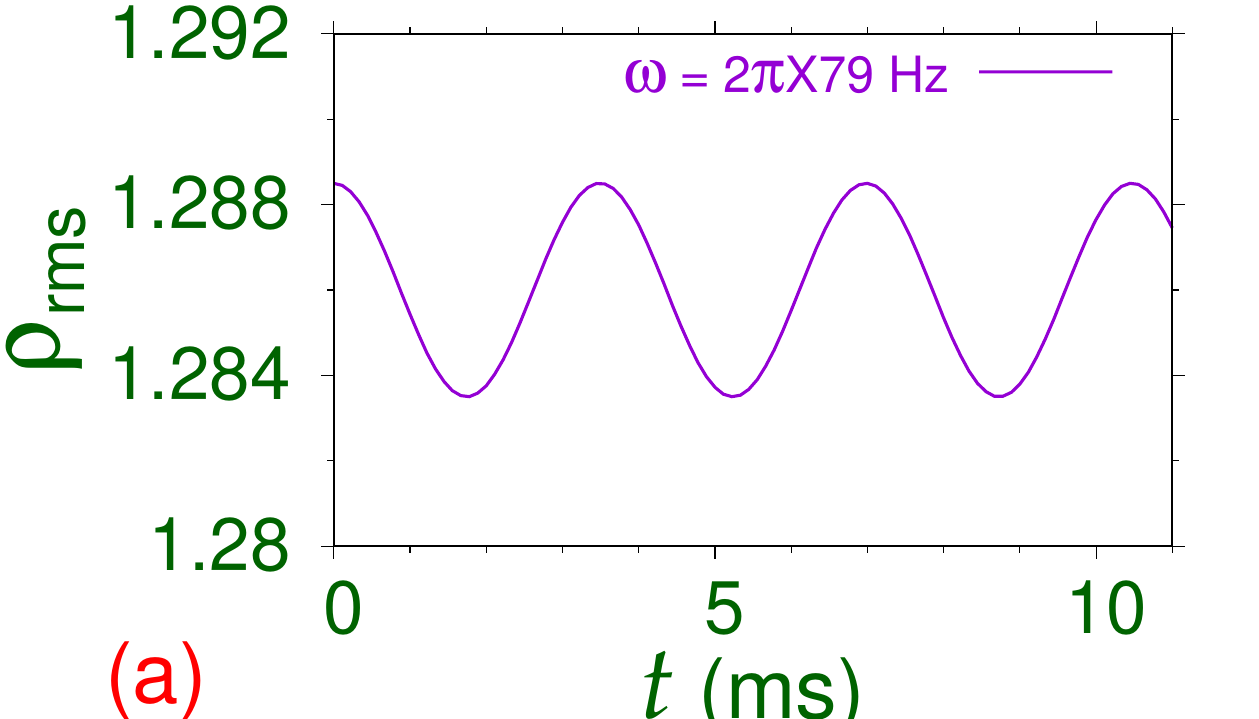}  
\includegraphics[width=.48\linewidth]{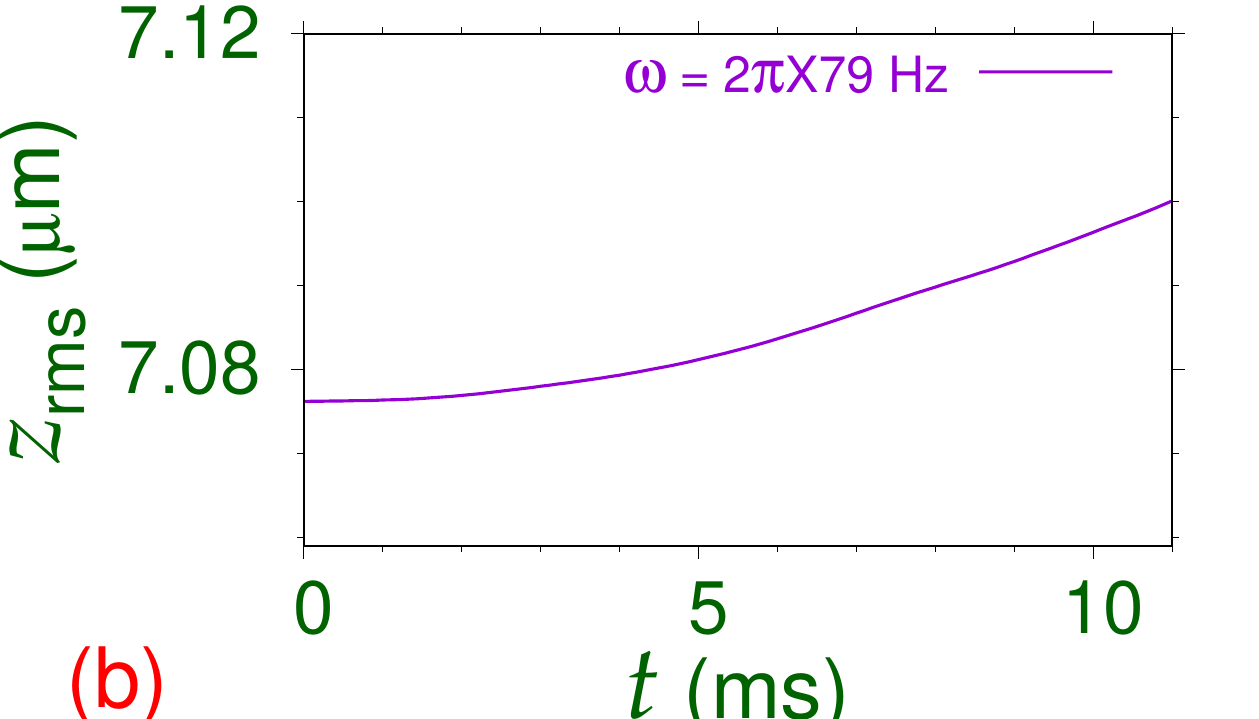}

\includegraphics[width=.6\linewidth]{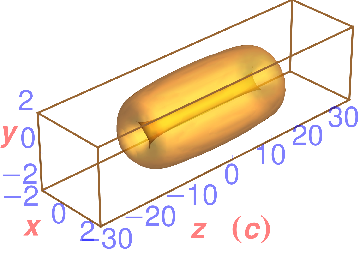} 
\includegraphics[width=.38\linewidth]{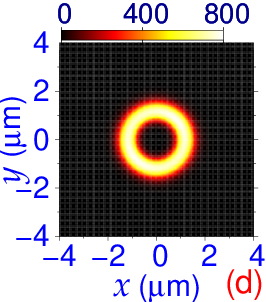}

\caption{Oscillation dynamics of  the hollow cylindrical droplet of Fig. \ref{fig2}(b) of $N=4000$ molecules 
as the angular frequency of the trap $\omega$ in Eq. (\ref{eq.GP3d}) is changed from $\omega =2\pi \times 78$ Hz to $2\pi \times 79$ Hz. 
 A plot of (a)   $\rho_{\mathrm{rms}}$ and (b) $z_{\mathrm{rms}}$ versus $t$ during this oscillation. 
 (c) Isodensity profile of  normalized density $|\psi(x,y,z)|^2$  of the droplet   at  $t=10$ ms, the density on the contour is $n_{\mathrm{contour}}  =4\times 10^{10}$ cm$^{-3}$.
 (d) Contour plot of integrated net 2D density $n(x,y)\equiv N n_{\mathrm{2D}}(x,y)$ in units of $\mu$m$^{-2}$
 versus $\{x,y\}$ of the droplet at $t=10$ ms.
All parameters are the same as in Fig. \ref{fig1}. 
}

\label{fig7} 
\end{center}
\end{figure}

As the hollow cylindrical droplets of ring topology are metastable states, they are only weakly stable. 
To demonstrate the  weak   stability of these droplets,   we perform two tests by real-time propagation after changing a parameter in Eq. (\ref{eq.GP3d}).
In the first test
we consider the converged wave function obtained by imaginary-time propagation of  the metastable droplet in Fig. \ref{fig2}(b)  with $N=4000$ molecules with the trap parameters of Fig. \ref{fig1} and with $a_{\mathrm{dd}}=2000a_0$ and $a=100a_0$. 
We investigate the   oscillation of this droplet by real-time propagation using the converged imaginary-time wave function as the initial state after changing the angular trap frequency of the harmonic trap in the $x$-$y$ plane from $\omega =2\pi \times 78$ Hz to $2\pi \times 79$ Hz. In Fig. \ref{fig7}(a) we illustrate the oscillation of the root mean square (rms)  
  radius of the droplet in the $x$-$y$ plane $\rho_{\mathrm{rms}}$
  through a plot of 
the same   versus time $t$. In  Fig. \ref{fig7}(b)  we plot the rms length of the droplet along the $z$ axis
$z_{\mathrm{rms}}$   versus time $t$ during the same oscillation.  In this case as the trapping frequency in the $x$-$y$ plane is increased, the oscillation starts with a reduction in $\rho_{\mathrm{rms}}$ and increase in $z_{\mathrm{rms}}$.    
Because of the radial confinement in the $x$-$y$ plane of the droplet, $\rho _{\mathrm{rms}}$ executes periodic oscillation over a reasonable period of time around the  rms value  $\rho_{\mathrm{rms}}$ in the final trap.  However, in the absence of any confining trap in the $z$ direction the rms length of the droplet in the $z$ direction slowly increases until a upper limit is attained.  The initial oscillation  of Fig.
\ref{fig7}(a)
is destroyed  at large times when the metastable hollow cylindrical droplet transforms eventually into  a stable one-droplet state (details not shown here).  In Fig. \ref{fig7}(c) 
we show the isodensity profile of the oscillating droplet 
and in Fig. \ref{fig7}(d) we present a contour plot of integrated net 2D density of the same droplet
at $t=10$ ms. 
 The hollow central region of the droplet is clearly visible in both plots  \ref{fig7}(c)-(d).   
 The periodic oscillation of    $\rho_{\mathrm{rms}}$ in  Fig. \ref{fig7}(a)  establishes the weak stability of the droplet.

 \begin{figure}[t!]
\begin{center}
\includegraphics[width=.48\linewidth]{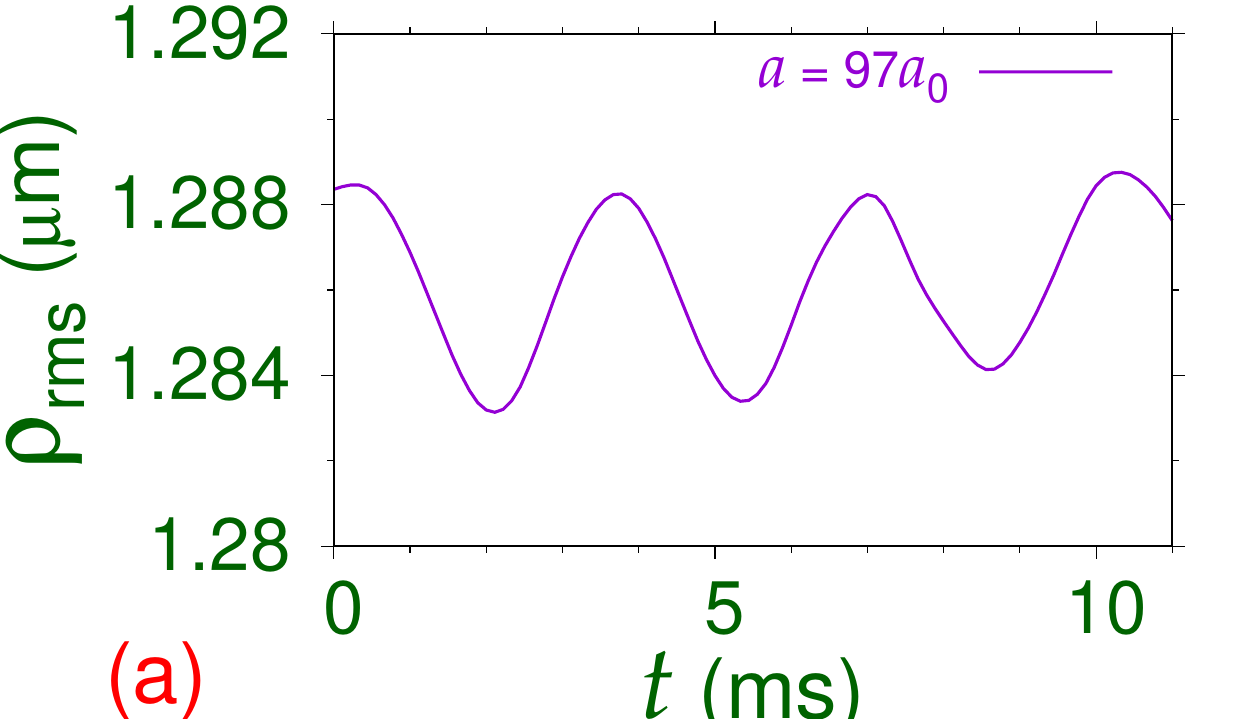} 
\includegraphics[width=.48\linewidth]{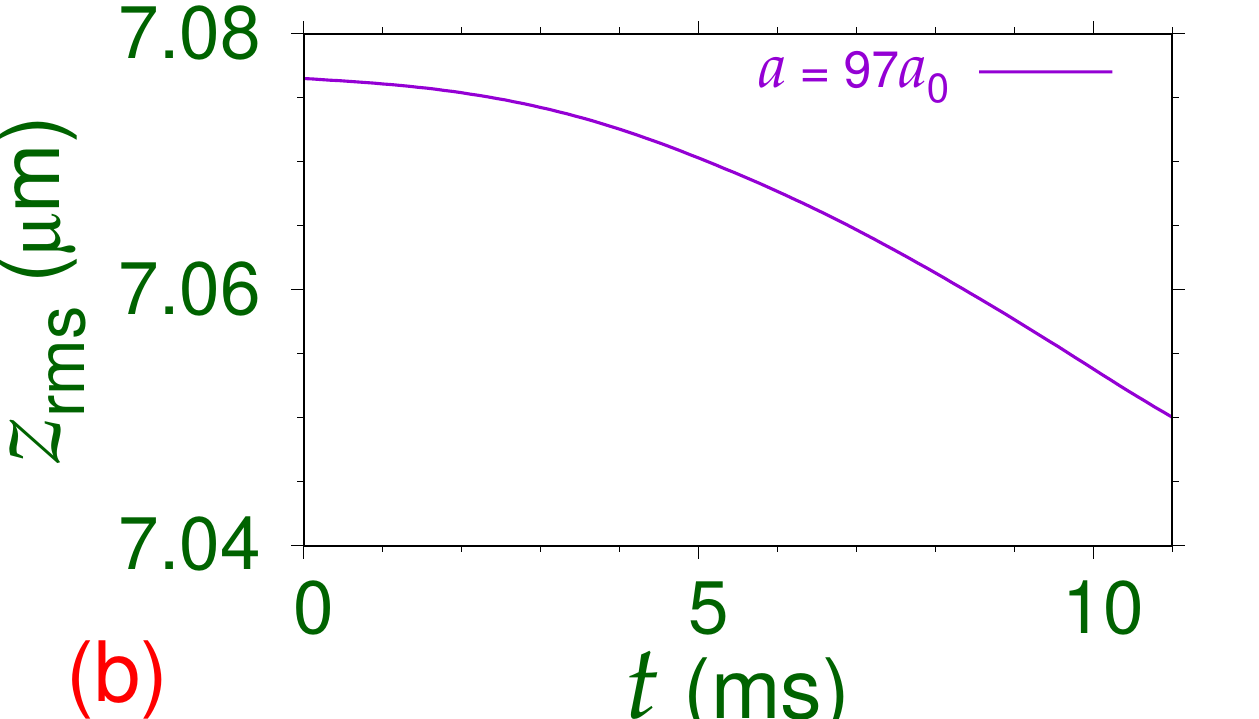}

\caption{Oscillation dynamics of  the hollow cylindrical droplet of Fig. \ref{fig2}(b) of $N=4000$ molecules
as the scattering length $a$   
 in Eq. (\ref{eq.GP3d}) is changed from $a = 100a_0$  to  $97a_0$. 
 A plot of (a)   $\rho_{\mathrm{rms}}$ and (b) $z_{\mathrm{rms}}$ versus $t$ during this oscillation. 
All parameters are the same as in Fig. \ref{fig1}. 
}

\label{fig8} 
\end{center}
\end{figure}

In the second test we study the real-time  dynamics by changing  the scattering length from $a=100a_0$ to $97a_0$.  In this case the contact repulsion is reduced which will enhance the dipolar attraction. Consequently, the system will be more bound  
and both  rms sizes   $\rho_{\mathrm{rms}}$ and $z_{\mathrm{rms}}$ will reduce initially. The resultant dynamics of the rms sizes is shown in Figs. \ref{fig8}(a)-(b) for $\rho_{\mathrm{rms}}$ and $z_{\mathrm{rms}}$. Again we find an oscillation of  $\rho_{\mathrm{rms}}$ during about 10 ms  and  after that, quite expectedly, the hollow cylindrical structure of the metastable droplet is destroyed and the droplet transforms into a one-droplet state. There being no trap in the $z$ direction the rms size $z_{\mathrm{rms}}$  reduces until a lower value is attained.
It is interesting to note that the frequency of the oscillation in plots \ref{fig7}(a) and \ref{fig8}(a) are very similar as this is controlled by the frequency $\omega$ of the harmonic trap in the $x$-$y$ plane.   The shape of the droplet at $t=10$ ms  is quite similar to the same in Figs. \ref{fig7}(c)-(d) (not shown in this paper).

 \begin{figure}[t!]
\begin{center}
\includegraphics[width=.32\linewidth]{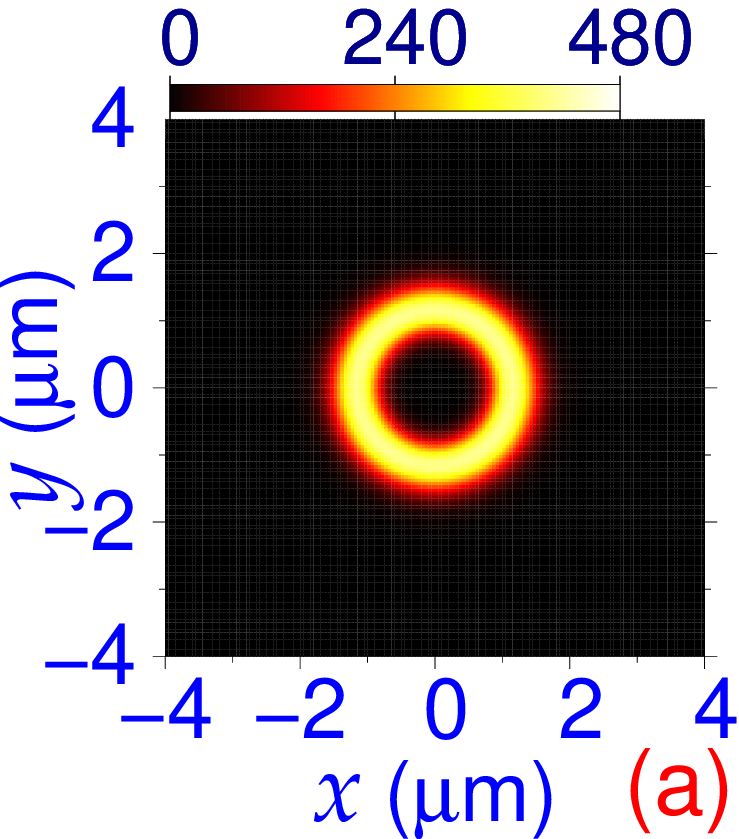} 
\includegraphics[width=.66\linewidth]{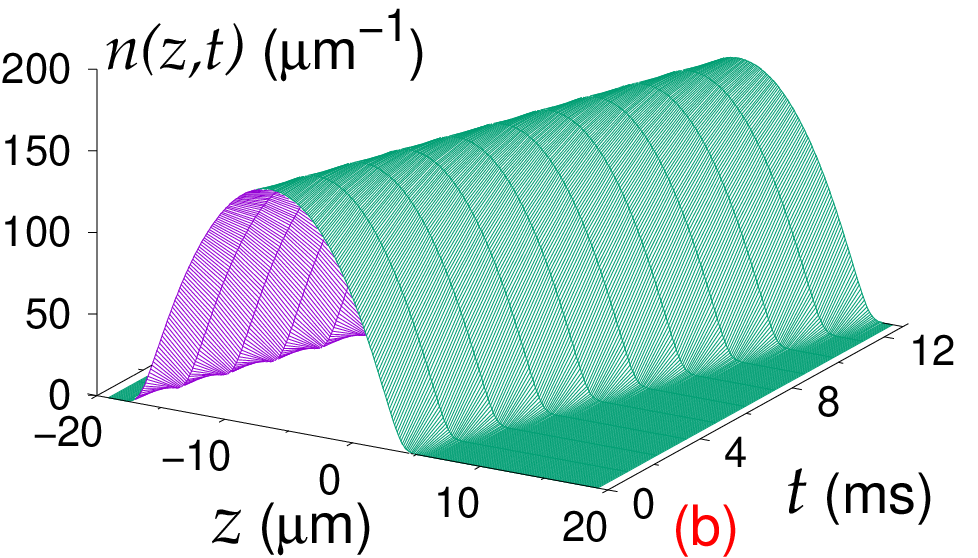}

\caption{Propagation dynamics of  the hollow cylindrical droplet of Fig. \ref{fig2}(a) of $N=2000$ molecules during time $t=0$ to 12 ms. In (a) we display a contour plot of net 2D density $n(x,y)$  versus \{x,y\} of this droplet at $t=12$ ms.  In (b) we show the time evolution of net integrated 1D density $n(z,t)$ during the uniform motion of the hollow cylindrical droplet.
All parameters are the same as in Fig. \ref{fig1}. 
}

\label{fig9} 
\end{center}
\end{figure}

To test the mobility of the hollow cylindrical droplets, we consider the converged imaginary-time wave function of the hollow cylindrical droplet  of  $N=2000$ molecules, viz.  Fig. \ref{fig2}(a),  multiply it by a phase factor $\exp(\mathrm{i}vz)$ and use the resultant function as the initial function in real-time propagation during an interval of time 12  ms.  
In  Fig. \ref{fig9}(a) we illustrate the contour plot of the net integrated 2D density $n(x,y) \equiv  Nn_{\mathrm{2D}}(x,y)$  versus $x,y$ of this droplet  as obtained by real-time propagation at $t=12$ ms.  We verified that in 12 ms  the hollow cylindrical droplet   moved 
from  -6.2 $\mu$m to  +6.2 $\mu$m  or a distance of $z_0=12.4$  $\mu$m resulting in a velocity of  about $v_0 =1$  mm/s. In Fig. \ref{fig9}(b) we show this propagation of the droplet through a plot of the net integrated 1D density $Nn(z,t)$ versus $z$ and $t$ during the uniform linear motion along the $z$ axis with the velocity of  about 1 mm/s. The hollow cylindrical profile is clearly visible in Fig. \ref{fig9}(a). This fact together with the smooth propagation of the  hollow cylindrical droplet in   Fig. \ref{fig9}(b) establishes the   mobility of the hollow cylindrical droplets.

\section{Summary}

\label{IV}  
 
 In this paper we have shown that, in a very strongly dipolar molecular condensate \cite{NaCs} with a very large dipolar length,
 it is possible to have a metastable  hollow-cylindrical quasi-1D droplet axially free along the polarization $z$ direction
 and subject to  an expulsive Gaussian potential  and a weak harmonic potential
$-$ both acting in the $x$-$y$ plane. In this demonstration we employ an improved mean-field model including a higher-order LHY repulsive interaction \cite{lhy}, appropriately modified for dipole-dipole interaction \cite{qf1,qf2}.
Specifically, in this study the dipolar length and the  scattering length of the  very strongly dipolar molecular condensate  are taken to be $a_{\mathrm{dd}}= 2000a_0$ and $a=100a_0$, and the number $N$ of molecules were kept between 2000 and 8000.
 The ground state of this system is the stable one-droplet state. In addition, there is a metastable two-droplet state as the first excited state.    The  hollow cylindrical   droplet is the second  excited state as illustrated in Fig. \ref{fig1}.  These quasi-1D  droplets  are localized in a small region  in the $x$-$y$ plane and elongated along the $z$ axis, like the droplets in a fully trapped dipolar BEC \cite{y13,2d3}, as shown in Figs. \ref{fig2}-\ref{fig5}.
 The repulsion in the $x$-$y$ plane  arising due to the combined action of the expulsive Gaussian potential and the dipole-dipole interaction, stabilizes the droplets. 
The GP model has a net cubic attractive nonlinear term arising from the contact and the dipole-dipole interaction and the higher-order   LHY interaction leads to a repulsive quartic nonlinearity, which stops the collapse of the strongly dipolar molecular  BEC to stabilize the droplets.  
 If the net dipole-dipole interaction of the system is reduced,  by reducing either the number of molecules or the dipolar length,
keeping all other parameters unchanged, the metastable hollow cylindrical  droplet  transforms  to  the stable  one-droplet state. 
We also demonstrated weak stability of this metastable droplet (as opposed to strong stability of ground-state stable droplets)  by real-time propagation after  changing  (a)  the trap frequency and (b) the scattering length by a small amount. 
This resulted in a steady oscillation of the droplet in the trapped $x$-$y$ plane during several cycles  demonstrating the weak stability.   
The present study will shed new lights on the generation of quantum states of nontrivial topology.


   
\section*{Acknowledgments}
SKA acknowledges support by the CNPq (Brazil) grant 301324/2019-0 and 303885/2024-6.


\begin{thebibliography}{99}
%



\bibitem{y13} M. Schmitt, M. Wenzel, F. Böttcher, I. Ferrier-Barbut, T. Pfau,
Nature 539, 259 (2016).
 \bibitem{2d3} 
 H. Kadau, M. Schmitt, M. Wenzel, C. Wink, T. Maier, I. Ferrier-Barbut, T. Pfau, Nature 530, 194 (2016).


\bibitem{santos}F. W\"achtler,  L. Santos, Phys. Rev. A 93, 061603(R) (2016).

\bibitem{drop3}F. W\"achtler, L.  Santos,   Phys. Rev. A 94, 043618 (2016).












\bibitem{blakie} D. Baillie, R. M. Wilson, R. N. Bisset, P. B. Blakie,
Phys. Rev. A 94, 021602(R) (2016).


\bibitem{c30} M. Edmonds, T. Bland, N. G. Parker, 	J. Phys. Commun. 4, 125008 (2020).

\bibitem{c31} S. De Palo, E. Orignac, R. Citro, Phys. Rev. B 106, 014503 (2022).


\bibitem{c32} E. Orignac, S. De Palo, L. Salasnich,
 R. Citro, Phys. Rev. A 109, 043316 (2024). 

\bibitem{c33}
Q. Zhu, 
C. Kong, Phys. Lett. A 536, 130291 (2025).


\bibitem{c27} R. Oldziejewski, W. Górecki, K. Pawlowski, K. Rzazewski,
Phys. Rev. Lett. 124, 090401  (2020).


\bibitem{NaCs}N. Bigagli, W. Yuan, S. Zhang, B. Bulatovic, T. Karman, I. Stevenson,  S. Will,
 Nature 631, 289 (2024).
 
 

 
 \bibitem{Bohn} G. Qu\'em\'ener, J. L. Bohn,  J. F. E.  Croft, 
 Phys. Rev. Lett. 131, 043402 (2023).





\bibitem{nat-gen} Md Shafayat  Hossain et al., Nature 628, 527  (2024).



\bibitem{ska}S. K. Adhikari,
Phys. Rev. A 85, 053631  (2012).




\bibitem{luis13}L. E. Young-S.,  S. K. Adhikari, Phys. Rev. A 108, 053323 (2023).








\bibitem{y15} L. Chomaz, D. Petter, P. Ilzhöfer, G. Natale, A. Trautmann,
C. Politi, G. Durastante, R. M. W. van Bijnen, A. Patscheider,
M. Sohmen, M. J. Mark, F. Ferlaino, Phys. Rev. X 9, 021012
(2019).
\bibitem{y16} G. Natale, R. M. W. van Bijnen, A. Patscheider, D. Petter, M.
J. Mark, L. Chomaz, F. Ferlaino, Phys. Rev. Lett. 123, 050402
(2019)
\bibitem{y17} L. Tanzi, E. Lucioni, F. Famà, J. Catani, A. Fioretti, C. Gabbanini, R. N. Bisset, L. Santos, G. Modugno, Phys. Rev. Lett. 122,
130405 (2019).
\bibitem{y18} F. Böttcher, J.-N. Schmidt, M. Wenzel, J. Hertkorn, M. Guo, T.
Langen, T. Pfau, Phys. Rev. X 9, 011051 (2019).




\bibitem{y19} M. Guo, F. Böttcher, J. Hertkorn, J.-N. Schmidt, M. Wenzel, H.
P. Büchler, T. Langen, T. Pfau, Nature 574, 386 (2019).








\bibitem{c2}R. M. Wilson, S. Ronen, J. L. Bohn,
Phys. Rev. A 80, 023614 (2009).

\bibitem{y6} T. Koch, T. Lahaye, J. Metz, B. Fröhlich, A. Griesmaier, T. Pfau,
Nature Phys. 4, 218 (2008).
\bibitem{c4} T. Lahaye, J. Metz, B. Frohlich, T. Koch, M. Meister,
A. Griesmaier, T. Pfau, H. Saito, Y. Kawaguchi, and
M. Ueda, Phys. Rev. Lett. 101, 080401 (2008).

\bibitem{c5} N. G. Parker, C. Ticknor, A. M. Martin, D. H. J.
O’Dell, Phys. Rev. A 79, 013617 (2009).
\bibitem{c6} J. L. Bohn, R. M. Wilson, S. Ronen, Laser Phys.
19, 547 (2009).


\bibitem{lhy}T. D. Lee, K. Huang, C. N. Yang, Phys. Rev. 106, 1135
(1957).

\bibitem{qf1}A. R. P.  Lima,  A. Pelster,   Phys. Rev. A 84, 041604(R) (2011).


\bibitem{qf2}{R. Sch\"utzhold, M. Uhlmann, Y. Xu, U. R. Fischer, Int. J. Mod. Phys. B 20,  3555 (2006).}
 
\bibitem{qf3} Z. Shi, G. Huang,
Phys. Rev. E 107, 024214 (2023). 

\bibitem{blakie1}D. Baillie, P. B. Blakie, Phys. Rev. Lett. 
121,195301 (2018).



\bibitem{luis23} L. E. Young-S., S. K. Adhikari,  Phys. Rev. A 105, 033311 (2022). 


\bibitem{santos1}E. Poli, T. Bland, C. Politi, L. Klaus, M. A. Norcia, F.
Ferlaino, R. N. Bisset, L. Santos, Phys. Rev. A 104,
063307 (2021).






\bibitem{pohl} Y.-C. Zhang, T. Pohl, F. Maucher, Phys. Rev. A 104,
013310 (2021).


\bibitem{rev} F. B\"ottcher, J.-N. Schmidt, J. Hertkorn, K. S. H. Ng,
S. D. Graham, M. Guo, T. Langen, T. Pfau,  Rep. Prog. Phys. 84,
012403 (2021).

\bibitem{rev2}{E. J. Halperin, S. Ronen, J. L. Bohn,
Phys. Rev. A 107, L041301 (2023). }

\bibitem{rev3} P. Ilzh\"ofer, M. Sohmen, G. Durastante, C. Politi,
A. Trautmann, G. Natale, G. Morpurgo, T. Giamarchi,
L. Chomaz, M. J. Mark, F. Ferlaino,  Nature Phys. 17, 356 (2021).


 \bibitem{ajp}S. K. Adhikari, Am. J. Phys. 54, 362  (1986).

 

\bibitem{dip} R. Kishor Kumar, L. E. Young-S., D. Vudragovi\'c, A. Bala\v{z}, P. Muruganandam, S. K. Adhikari, Comput. Phys. Commun. { 195}, 117 (2015).





\bibitem{yuka}V. I. Yukalov, Laser Phys.  28, 053001 (2018).



\bibitem{dipbec} T. Lahaye, C. Menotti, L. Santos,  M. Lewenstein,  T. Pfau, Rep.
Prog. Phys. 72, 126401 (2009).

\bibitem{dipbecx}L. Chomaz, I. Ferrier-Barbut, F. Ferlaino,
B. Laburthe-Tolra, B. L. Lev, T. Pfau, Rep.
Prog. Phys. 86, 026401 (2023).










  
  
  
  
  
  
  
  
  
  
  
  
  
  
 \bibitem{rmp}F. Dalfovo, S. Giorgini, L. P. Pitaevskii, S. Stringari,
Rev. Mod. Phys. 71, 463 (1999).
 

\bibitem{yuka1}P. B. Blakie, C. Ticknor, A. S. Bradley, A. M. Martin, M. J. Davis, Y. Kawaguchi,
Phys. Rev. E 80, 016703 (2009).

\bibitem{omp}V. Lon\v car, L. E. Young-S., S. \v Skrbi\'c, P. Muruganandam, S. K. Adhikari, A. Bala\v z, Comput. Phys. Commun. { 209}, 190 (2016).

\bibitem{ompF} L. E. Young-S., P. Muruganandam, A. Bala\v z, S. K. Adhikari,
Comput. Phys. Commun.  286, 108669 (2023).

\bibitem{crank}P. Muruganandam, S. K. Adhikari, Comput. Phys. Commun. 180, 1888       (2009).


 \bibitem{58}F. Deng, X.-Y. Chen, X.-Y. Luo, W. Zhang, S. Yi, T. Shi,  Phys. Rev. Lett. 130 183001
(2023).

  

\bibitem{dm1} B. Xu, F. Yang, R.  Qi, H.  Zhai,  P.  Zhang,  arXiv:2410.10806  (2024).
 
\bibitem{dm2} T. Langen, J. Boronat,  J. Sánchez-Baena, R. Bombín, T. Karman,  F. Mazzanti, 
Phys. Rev. Lett. 134, 053001  (2025).


\bibitem{aa} A. Cidrim, F. E. A. dos Santos, E. A. L. Henn,  T. Macrì,
Phys. Rev. A 98, 023618 (2018).

\bibitem{b}G. Li, Z. Zhao, X. Jiang, Z. Chen, B. Liu, B. A. Malomed,  Y. Li,
Phys. Rev. Lett. 133, 053804 (2024).

\bibitem{c}G. Li, X. Jiang, B. Liu, Z. Chen, B. A. Malomed, Y. Li,
Front. Phys. 19, 22202 (2024).

\bibitem{d}E. A. L. Henn,   Front. Phys. 19, 31301 (2024).


 
 
\end{thebibliography}
\end{document}